\documentclass[preprint,12pt]{aastex}

\renewcommand\th{\thinspace}
\newcommand\kms{\ifmmode{\rm km\th s^{-1}}\else km\th s$^{-1}$\fi}
\newcommand\cmss{\ifmmode{\rm cm\th s^{-1}}\else cm\th s$^{-2}$\fi}
\newcommand\msun{\ifmmode{M_{\odot}}\else $M_{\odot}$\fi}
\newcommand\rsun{\ifmmode{R_{\odot}}\else $R_{\odot}$\fi}

\begin{document}
\title{A Dynamical Mass Constraint for Pre-Main-Sequence Evolutionary Tracks: The Binary NTT 045251+3016}
\author{Aaron T. Steffen, Robert D. Mathieu}
\affil{Department of Astronomy, University of Wisconsin--Madison, 475 
  North Charter Street, Madison, WI 53706-1582}
\email{steffen@astro.wisc.edu; mathieu@astro.wisc.edu}
\author{Mario G. Lattanzi}
\affil{Osservatorio Astronomico di Torino, Strada Osservatorio 20, I-10025 Pino Torinese, Italy}
\email{lattanzi@gsc2.to.astro.it}
\author{David W. Latham}
\affil{Harvard-Smithsonian Center for Astrophysics, 60 Garden Street, Cambridge, MA  02138}
\email{dlatham@cfa.harvard.edu}
\author{Tsevi Mazeh}
\affil{School of Physics and Astronomy, Raymond and Beverly Sackler Faculty of Exact Sciences, Tel Aviv University, Tel Aviv 69978, Israel}
\email{mazeh@wise.tau.ac.il}
\author{L. Prato\altaffilmark{1}}
\affil{Department of Physics and Astronomy, UCLA, Los Angeles, CA 90095-1562}
\email{lprato@astro.ucla.edu}
\author{Michal Simon\altaffilmark{1}}
\affil{Department of Physics and Astronomy, SUNY, Stony Brook, NY 11794-3800}
\email{MSIMON@astro.sunysb.edu}
\author{Hans Zinnecker}
\affil{Astrophysikalisches Institut Potsdam, An der Sternwarte 16, D-14482 Potsdam, Germany}
\email{hzinnecker@aip.de}
\author{Davide Loreggia}
\affil{Osservatorio Astronomico di Torino, Strada Osservatorio 20, I-10025 Pino Torinese, Italy}
\email{loreggia@to.astro.it}

\altaffiltext{1}{Visiting Astronomer, Kitt Peak National Observatory,
National Optical Astronomy Observatories, operated by Association of
Universities for Research in Astronomy, under cooperative agreement.}

% Here is the abstract
\begin{abstract}
We present an astrometric/spectroscopic orbital solution for the
pre-main-sequence binary NTT 045251+3016.  Interferometric
observations with the HST FGS3 allowed stellar separations as small as
14 mas to be measured.  Optical spectra provided 58 radial-velocity
measurements of the primary star and near-infrared spectra provided 2
radial-velocity measurements of both the primary and secondary, giving
a mass ratio for the binary system.  The combination of these data
allows the dynamical masses and the distance of the stars to be
derived.  Our measurements for the primary and secondary masses are
1.45 $\pm$ 0.19 M$_\odot$ and 0.81 $\pm$ 0.09 M$_\odot$, respectively,
and 145 $\pm$ 8 pc for the distance of the system, consistent with
prior estimates for the Taurus-Auriga star-forming region.  The
evolutionary tracks of D'Antona \& Mazzitelli (1997), Baraffe et
al. (1998), and Palla \& Stahler (1999) are tested against these
dynamical mass measurements.  Due to the intrinsic color/$T_{eff}$
variation within the K5 spectral class, each pre-main-sequence model
provides a mass range for the primary. The theoretical mass range
derived from the Baraffe et al. (1998) tracks that use a mixing length
parameter $\alpha = 1.0$ is closest to our measured primary mass,
deviating between 1.3 and 1.6 sigma.  The set of Baraffe et al. (1998)
tracks that use $\alpha = 1.9$ deviate between 1.6 and 2.1 sigma from
our measured primary mass.  The mass range given by the Palla \&
Stahler (1999) tracks for the primary star deviate between 1.6 and 2.9
sigma.  The D'Antona \& Mazzitelli (1997) tracks give a mass range
that deviates by at least 3.0 sigma from our derived primary mass,
strongly suggesting that these tracks are inconsistent with our
observation.  Observations of the secondary are less constraining than
those of the primary, but the deviations between the dynamical mass of
the secondary and the mass inferred for the secondary from the various
pre-main-sequence tracks mirror the deviations of the primary star.
All of the pre-main-sequence tracks are consistent with coevality of
the components of NTT 045251+3016.
\end{abstract}


\section{Introduction}
Pre-main-sequence (PMS) stellar evolution has been extensively modeled
in the last decade.  Unfortunately, very few PMS stars have provided
detailed observational tests of these models, in large part due to the
lack of dynamical mass determinations. Recently a precise mass
determination has been achieved for the secondary of the eclipsing
binary TY~CrA (Casey et al.\ 1998, Corporon et al.\ 1996, and
references therein). Casey et al.\ tested three sets of evolutionary
tracks against the 1.64 M$_{\odot}$ secondary, and found that all were
consistent with the observed physical parameters.  Covino et
al. (2000) analysed the eclipsing binary RXJ 0529.4+0041 and compared
the 1.25 M$_{\odot}$ primary star and the 0.91 M$_{\odot}$ secondary
star with three sets of evolutionary tracks. They found that the
Baraffe et al. (1998) tracks provide the closest agreement with the
derived masses of both stars. Stellar masses have also been measured
via orbital motions of disk gas (e.g., Guilloteau et al.\ 1999, Simon
et al.\ 2001).  This is currently the only technique available to
obtain the mass of a single star, but it is limited by distance
uncertainty.  Simon et al. (2001) found that PMS evolutionary models
that presented cooler T$_{eff}$ (e.g. Baraffe et al.\ 1998 and Palla
\& Stahler 1999) provided a better fit with their derived masses of
nine PMS stars.  Another powerful method for stellar mass
determination is the combination of spectroscopic and astrometric
observations of a binary star.  In addition to all of the orbital
elements, the combination of an angular measure of the orbit from
astrometry and a linear measure from radial velocities allows an
independent determination of the binary distance.  This paper
describes such a study of the naked T Tauri binary NTT 045251+3016.

	NTT 045251+3016 was first discovered with the Einstein X-ray
Observatory as an optically visible star associated with an X-ray
source.  Detailed UBVRIJHKL photometry and optical spectra for the
primary star are presented by Walter et al.~(1988), who derived a K7
spectral classification for the primary star, and also identified the
system as a spectroscopic binary.  Improved spectral classification of
the primary is discussed in Section 4.1.  Based on a mean radial
velocity near to that of the Taurus-Auriga association and a Li~I
$\lambda$6707\AA\ equivalent width of 0.58\AA\, Walter et al. (1988)
identified NTT 045251+3016 as a PMS star.

	NTT 045251+3016 is unique among known PMS binaries in that it
has an orbit with a short enough period to permit measurable
radial-velocity variations of the primary and secondary stars while at
the same time being wide enough to spatially resolve the two stars
with HST's Fine Guidance Sensor. We describe here an observational
program to obtain both spectroscopic and astrometric data, thereby
determining the masses of both PMS stars and the distance to the
binary.

\section{Observations}
\subsection{Astrometric}
  The first interferometric observation of NTT 045251+3016 with the
astrometer Fine Guidance Sensor No.3 (FGS3) aboard HST was executed on
April 14, 1995. The target was visited on 17 different occasions
during the subsequent 3.3-year observing campaign (completed Aug 20,
1998), covering approximately one quadrant of the apparent orbit.  The
interferometric mode of FGS3 is used to sample the visibility function
(VF) produced by the Koester's prism interferometer as the
field-of-view of the unit is driven across the target. For best
results, the FGS was commanded to oversample the VF by taking
measurements every 0.6 mas (on the sky) during a scan of $\sim$ 1500
mas in length. The scan length used is sufficient, during normal
operations, to encompass the sensitivity range of the interferometer,
which extends for $\sim$ 20 mas around the line-of-sight to the
target.  FGS3 is endowed with two Koester's prisms that provide
sensitivity in two orthogonal directions, usually referred to as the
FGS X and Y axes (Lupie \& Nelan 1998). Therefore, each scan of the
target results in two VF's which are then independently analyzed for
signatures other than those characteristic of a suitable single (and
point-like) star chosen as template.  Deviations from the single-star
VF provide measures of the projected separations of the binary at each
observing epoch. The comparison with the template also provides two
independent estimates of the magnitude difference between the two
companions. On-sky separation ($\rho$) and astronomical position angle
(PA) are easily derived from the projected separations and telescope
attitude data (Bernacca et al. 1993).

Fifteen consecutive scans were taken on each visit to check on
scan-to-scan repeatability and for improvement of the S/N. The average
S/N ratio of the observed scans is $\sim$ 20, which improved to $\sim$
$20~ \times ~ \sqrt{15} ~ \sim ~ 80$ after merging of the scans
(Lattanzi et al. 1997, and references therein).  The shape of the
visibility curve changes with effective wavelength, which can
introduce systematic errors if the color of the template differs
significantly from that of the target. To minimize this effect, we
selected star SAO185689 ((B-V)=1.5 mag) as the template single
star. Among the templates made available by the ST ScI for FGS
reductions, this is the star with the color closest to that of our
binary (observed (B-V)=1.28 mag).

Fourteen of the 17 visits produced successful measurements of the
binary separation and orientation. These measurements are listed in
Table 1. The 85\% success rate is gratifying given both the small
separation ($\sim$ 3 times smaller than the Airy angular resolution
limit at visible light) and the relatively large magnitude difference
(more than 2 magnitudes).

The formal error of the projected separations is $\sim$ 1 mas on each
axis.  However, this error is internal, as it only takes into account
the contribution due to the reduction method, which is based on an
analytical cross-correlation technique (Bernacca et al. 1993).  There
are other error sources which are known to be present but are
difficult to quantify for each observation. As an example, the epochs
of observation of the calibration star are usually quite different (up
to 5 months) from the dates of the visits to our target; consequently,
the template star might not be the best representation of the FGS
signatures at the time of the science observation, as small changes
are known to occur even on relatively short time scales. By comparing
observations of the template star taken at different times, we have
measured variations in the structure of the VF which could increase
the separation measurement error to 2-3 mas. Therefore, this is
probably a more realistic range for the error of several of the
measured projected separations in Table 1. This corresponds to 3-4 mas
for the (1 $\sigma$) error on total separation and to $\sim$
14$^\circ$ for the maximum error in PA.

Table 1 also shows the measurements of the visual magnitude difference
between the two companions. The values derived from the independent
fits to the X and Y VF's are generally in good agreement except for
the tenth and twelfth measurements. Averaging the results yields
$\Delta m_X \simeq$ 2.2 $\pm$ 0.3 mag and $\Delta m_Y \simeq$ 2.3
$\pm$ 0.3 mag for the average magnitude differences on the X and Y
axes, respectively. Again, these results are consistent with
instrument performances expected in a challenging scenario like ours,
i.e., small separation and relatively large magnitude difference.

\subsection{Optical Radial Velocities}
Since 1985 we have monitored the radial velocity of NTT 045251+3016
with the Center for Astrophysics (CfA) Digital Speedometers (Latham
1992).  Three nearly identical instruments were used on the Multiple
Mirror Telescope and 1.5-m Tillinghast Reflector at the Whipple
Observatory atop Mt. Hopkins, Arizona, and on the 1.5-m Wyeth
Reflector located in Harvard, Massachusetts.  Echelle spectrographs
were used with intensified photon-counting Reticon detectors to record
about 45~\AA\ of spectrum in a single order centered near 5187~\AA,
with a resolution of about 8.3~\kms\ and signal-to-noise ratios
ranging from 8 to 15 per resolution element.

Radial velocities were derived from the 58 observed optical spectra
using the one-dimensional correlation package {\bf rvsao} (Kurtz \&
Mink 1998) running inside the IRAF\footnotemark \footnotetext{IRAF
(Image Reduction and Analysis Facility) is distributed by the National
Optical Astronomy Observatories, which are operated by the Association
of Universities for Research in Astronomy, Inc., under contract with
the National Science Foundation.} environment.  The template spectrum
was drawn from a new grid of synthetic spectra (Morse \& Kurucz, in
preparation) calculated using model atmospheres computed using
Kurucz's code ATLAS9.  We correlated our observed spectra against a
grid of solar metallicity templates and chose the one that gave the
highest average peak correlation.

The highest peak correlation averaged over all 58 observed spectra was
obtained for $T_{\rm eff} = 4500$ K, $\log g = 3.5$ \cmss, $[m/H] =
0.0$, and $v \sin i = 10$ $\kms$.  The heliocentric velocities derived
with this template are reported in Table 2, together with the
heliocentric Julian date and (O-C) errors.  The rms deviation about
the orbital solution is 0.7 $\kms$, typical of CfA precision for
late-type dwarfs.  An implicit assumption of our analysis is that the
spectra of PMS stars can be reliably modeled using normal stellar
atmospheres.  However, a small template mismatch would not affect
significantly the accuracy of the radial-velocity measurements.

We attempted to detect the secondary spectrum in the optical with high
S/N observations of NTT 045251+3016 taken in February 1992, near a
time of maximum velocity separation, with the Hamilton echelle
spectrograph on the 3-m Shane telescope of Lick Observatory.  The
wavelength region covered was from 3800\AA\ to 9800\AA\ with a S/N of
30.  Cross-correlating our high signal-to-noise spectrum with a
variety of templates did not reveal the secondary's spectrum.  To test
our detection limits, C. Dolan created synthetic double-lined spectra
using a narrow-lined K6V spectral standard rotationally broadened to
10 $ \kms$ to match the rotationally broadened spectrum of NTT
045251+3016.  This primary spectrum was combined with numerous
secondary spectra ranging in spectral type (K2V to M2V), in rotational
velocities (5 to 50 $\kms$), in mass ratios (1.0 to 0.5), and in flux
ratios (1.0 to 0.05).  Dolan found that to detect the secondary's
spectrum a minimum flux ratio of $\sim$ 0.2 - 0.4 is required, except
in cases with extreme rotational broadening ($>40$ $ \kms$) when the
secondary becomes undetectable even with a luminosity ratio of 1.0.
Thus our non-detection is consistent with our FGS determined magnitude
difference of 2.4 mag in \emph{V}, corresponding to a flux ratio of
0.11.

\subsection{Near-Infrared Radial Velocities} 

Since we were unable to detect the secondary in the optical, we
obtained two spectra of the binary with PHOENIX, the KPNO
high-resolution near-infrared spectrograph on the 4-m Mayall
telescope.  Details on this work are given in Mazeh et al. (2000a,b).

The central wavelength of the observations was 1.555$\mu$m and yielded
a free spectral range of $\sim 1450$ $\kms$; the effective resolution
was 35,000.  We also observed a sample of 12 main-sequence
spectral-type standards from F6 through M7, in order to provide
templates for a two-dimensional cross-correlation analysis using
TODCOR (Zucker \& Mazeh 1994). We successfully detected the secondary
spectrum using spectra of HR 8085 (61 Cyg A; K5) and HR 8086 (61 Cyg
B; K7), each rotationally broadened by $15 ~ \kms$, as the primary and
secondary templates, respectively. The velocities are given in Table
3.  Our derived flux ratio at 1.555$\mu$m is $0.4 \pm 0.1$.

To estimate our uncertainty due to template mismatch we used TODCOR to
derive the velocity of each template with regard to the other
templates.  The scatter of these derived velocities is less than $0.7$
$\kms$. This error is added in quadrature to the estimated uncertainty
of the peak location of each spectrum, for a total uncertainty of
$0.9$ $\kms$.  The radial velocities of the primary derived from these
NIR spectra are consistent to within this uncertainty with the radial
velocities predicted by the primary orbital solution at the same
epoch, which suggests that our radial-velocity zero-point is also
accurate at this level.

\begin{figure}[t]
\plotone{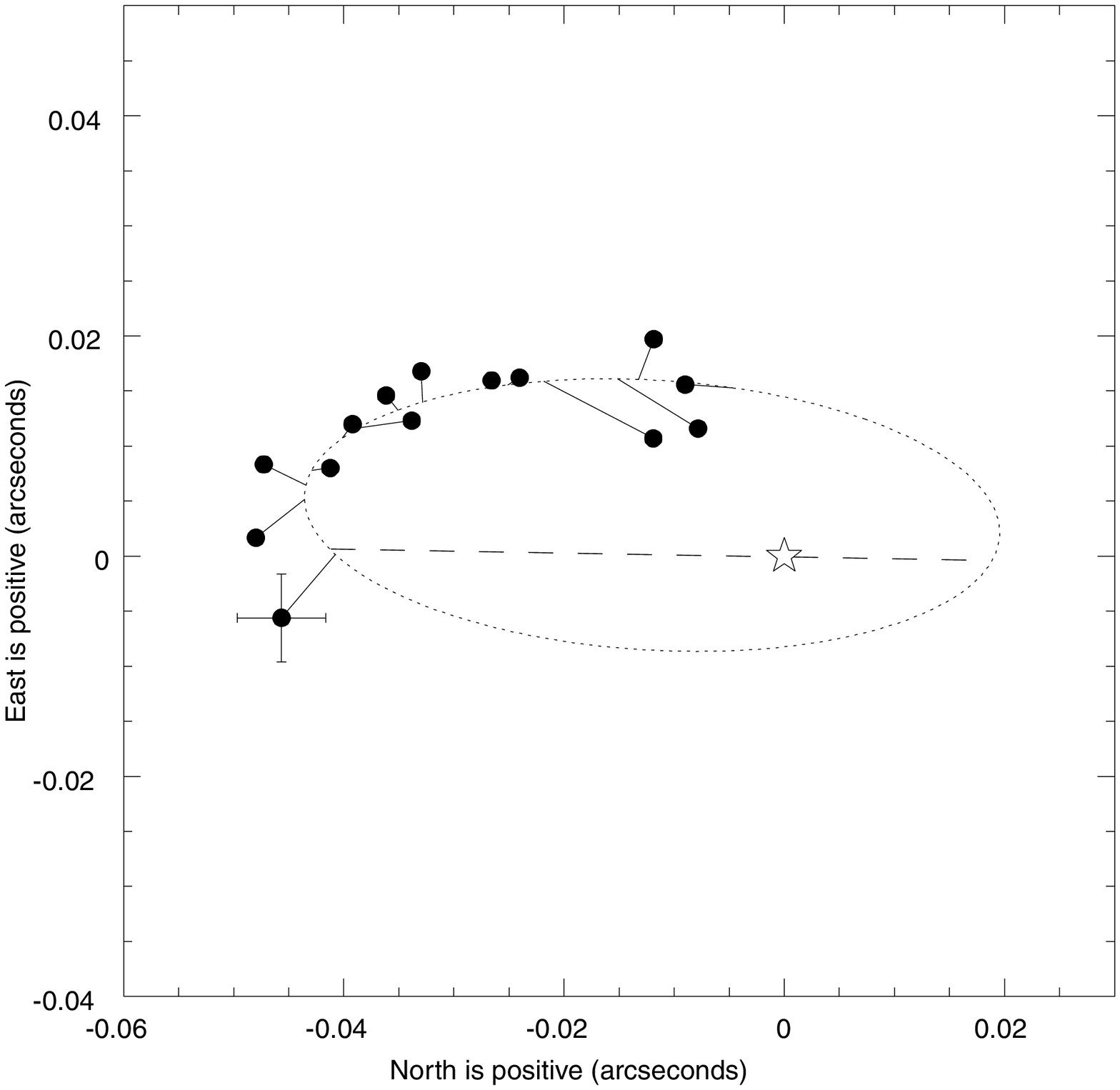}
\caption{Relative astrometric orbit. The filled circles are HST FGS
measurements, with lines indicating the predicted positions from
orbital solution.  4 mas error bars are shown on the first
observation. The star symbol shows the position of the primary
star. The dashed line represents the line of nodes.}
\label{fig:1}
\end{figure}

\begin{figure}[t]
\plotone{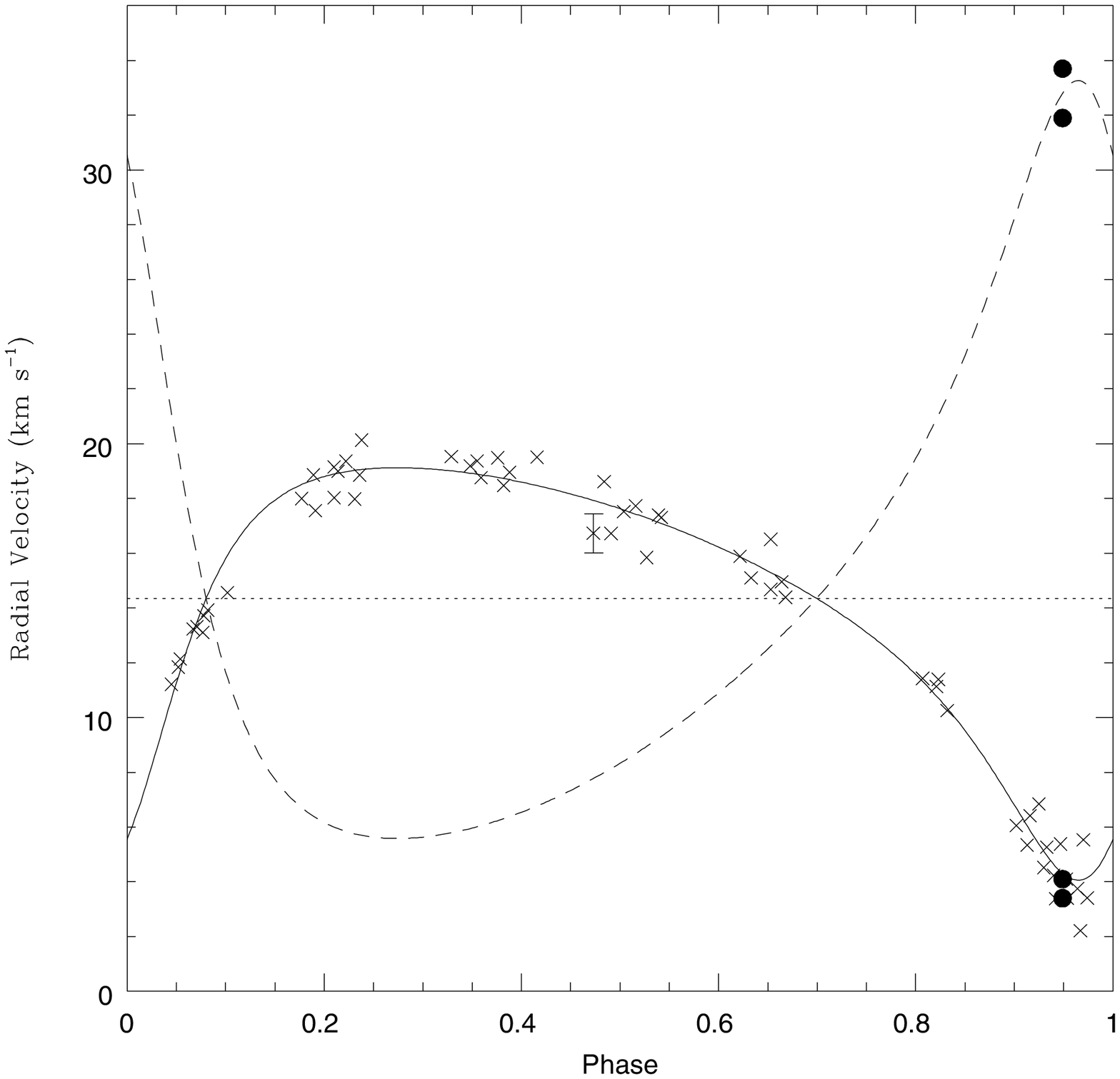}
\caption{Primary (solid) and secondary (dashed) spectroscopic orbit
solutions and observed radial velocities.  Optical data are presented
as X's and near-infrared data are solid circles.  The dotted line is
the center-of-mass ($\gamma$) velocity of the orbital solution.  0.7
$\kms$ error bars appear on a central point.}
\label{fig:2}
\end{figure}

\section{Orbital Solution}

Nine of the relative orbital elements of NTT 045251+3016 were
calculated via a simultaneous fit to the relative astrometric and
primary radial-velocity data, using a program generously provided by
G. Torres.  Table 4 shows the orbital elements of the binary NTT
045251+3016.  The astrometric and spectroscopic data and orbital
solutions are presented in Figures 1 and 2, respectively.  Although
the orbital solution is solved via a simultaneous fit to both the
relative astrometric and primary radial-velocity data, the large
number of radial-velocity measurements provide a strong constraint on
the elements of the orbital solution that can be calculated with
single-lined spectroscopic data (i.e.  $P$, $\gamma$, $K_{1}$, $e$,
$\omega$, $T$).  Thus, in practice, the orbital parameters that are
being determined with the astrometric data are the angular size of the
semi-major axis (\emph{a}($''$)), the position angle of the line of
nodes ($\Omega_{2000}$), and the inclination angle (\emph{i}).

Furthermore, the only parameter influencing the mass determination
which is derived from the astrometric data is the inclination.  As can
be seen from the orientation of the line of nodes in Figure 1, the
inclination is quite well constrained by the phase coverage of these
data.  The distance to the system, essentially set by the angular
semi-major axis, is more sensitive to the partial phase coverage and
possible systematic errors in the FGS astrometry.

With detection of the secondary in the near-infrared spectra, the mass
ratio, \emph{q}, was calculated using the \emph{calculated} radial
velocity of the primary from the orbital solution at the time of the
near-infrared observations, the center-of-mass velocity of the binary,
and the mean of the two \emph{measured} radial velocities of the
secondary which were obtained from IR spectra.  This mass ratio, in
combination with the 9 known orbital elements, allowed us to calculate
the individual masses of the stars in the binary and the distance to
the system.

The masses of the primary and secondary stars are given by the equations (Batten 1973): 

\begin{equation}
M_{1} \, [M_{\odot}] = \frac{3.793 \times 10^{-5} \, \sqrt[3]{(1-e^2)} \: (K_{1}+K_{1}/q)^{2} \: (K_{1}/q) \: P}{(\sin{i})^{3}}
\end{equation}

\begin{equation}
M_{2} \, [M_{\odot}] = \frac{3.793 \times 10^{-5} \, \sqrt[3]{(1-e^2)} \: (K_{1}+K_{1}/q)^{2} \: K_{1} \: P}{(\sin{i})^{3}}
\end{equation}
where $P$ is in years and $K_{1}$ is in km s$^{-1}$.  

We determine the masses of the primary and secondary to be $1.45 \pm
0.19$ M$_{\odot}$ and $0.81 \pm 0.09$ M$_{\odot}$, respectively.  From
Equation 1 we find that the largest contributors to the variance of
the dynamical mass of the primary are the measurement error associated
with the radial velocity of the secondary ($39\%$ of the variance) and the error on
the measurement of the orbital inclination ($38\%$).  The largest
contributer to the variance of the secondary's dynamical mass comes
from the error associated with the orbital inclination of the binary
($54\%$), followed by the error associated with the measured radial
velocity of the secondary ($17\%$). We note that the mass
uncertainties associated with our limited number of secondary
radial-velocity measurements can be reduced substantially now with
additional observations; higher precision astrometry awaits future
instrumentation.

The dynamical distance \emph{d} in parsecs is given by:

\begin{equation}
d = \frac{0.03357 \sqrt{(1-e^2)} \: (K_{1}+K_{1}/q) \: P}{a \: \sin{i}}
\end{equation}
where $a$ is in arcseconds. Our direct distance measurement to the PMS
binary of 145 $\pm$ 8 pc agrees with prior distance estimates to the
Taurus-Auriga complex based on indirect methods (e.g., 140 $\pm$ 10 pc
(Kenyon et al. 1994)) and geometric parallax (e.g., $139^{+10}_{-9}$
pc (Bertout et al.~1999), and $142 \pm 14$ pc (Wichmann et al. 1998)).
The center-of-mass velocity of the system, $14.35 \pm 0.11$, is
somewhat lower than the mean radial velocity of the Taurus-Auriga
complex (17.4 \kms; Hartmann et al.~1986), but with an association
velocity dispersion of 2 $\kms$ the binary radial velocity is
consistent with membership in the association. In this context we also
note that the binary is located at the edge of the association in the
Auriga subcomplex.

The largest contributor to the variance of the distance to NTT
045251+3016 comes from the error on the measurement of the angular
semi-major axis ($50\%$).  Since our astrometric data cover only a
little more than one quadrant of the orbit, it does not provide a
tight constraint on the measurement of the angular semi-major axis
(see Figure 1).

Table 4 summarizes our measured masses and distance to the components
of NTT 045251+3016.

\section{Comparison with PMS Evolutionary Models}
There are now a variety of PMS models that provide the luminosity and
effective temperature of a star given its mass and age. The
differences between these models lie largely in the choice of
opacities, atmospheres, metallicities, and convection models.  In
principle, a PMS star of known mass, luminosity, effective
temperature, and metallicity can distinguish between these differences
in stellar physics. NTT 045251+3016 provides such a case.

\subsection{The Primary Star}

The primary star of NTT 045251+3016 was originally classified as a K7
spectral type (Walter et al.~1988).  However, analysis of eight
temperature sensitive lines used for spectral classification by Lee
(1992) (VI $\lambda$6040\AA\, FeI $\lambda$6042\AA\, FeI,
$\lambda$6056\AA\, VI $\lambda$6058\AA\, NiI $\lambda$6108\AA\, VI
$\lambda$6112\AA\, FeI $\lambda$6200\AA\, ScI $\lambda$6211\AA ) in
our Hamilton echelle spectra clearly point to a K5 classification with
an uncertainty of less than one subtype.  The FeI and ScI line pair
around $\lambda$6200\AA\ in particular was used by Basri \& Batalha
(1990) for spectral classification.  Figure 3 compares this line pair
from our spectra with three spectral standards HR 8832 (K3V), HR 8085
(K5V), and HR 8086 (K7V).

\begin{figure}[t]
\epsscale{0.50}
\plotone{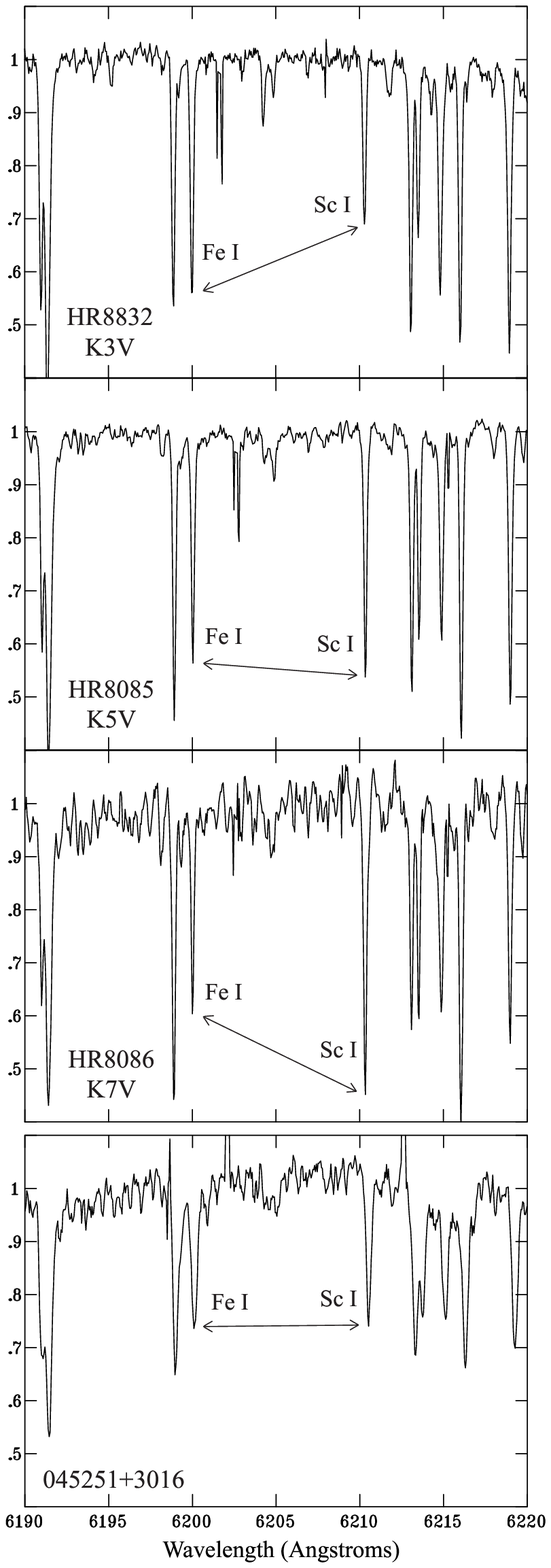}
\caption{FeI $\lambda$6200\AA\ and ScI $\lambda$6211\AA\ line pair
used for spectral classification. Spectra are shown for K3, K5, and K7
spectral type standards along with NTT 045251+3016.  A
three-point-smooth was performed on the spectrum of NTT 045251+3016
for presentation purposes.}
\label{fig:3}
\end{figure}

Using the photometry of Walter et al. (1988), the luminosity ratios in
\emph{V} and \emph{H}, and the intrinsic colors, temperature
calibrations, and bolometric corrections from Kenyon \& Hartmann
(1995), we find for the primary star an effective temperature of 4350
K and a luminosity of 0.75 L$_{\odot}$.  Since we do not have a
luminosity ratio in \emph{B}, and thus could not calculate the
\emph{B} magnitude for the primary and secondary individually, we
calculated the extinction in \emph{V} using two different methods.
First we assumed that the secondary did not contribute to the light in
the \emph{B} band, so the \emph{B} magnitude calculated in Walter et
al. (1988) (B=12.88) was the \emph{B} magnitude of the primary.  Using
this observed \emph{B} magnitude and the calculated \emph{V} magnitude
of the primary with the \emph{(B-V)} color for a K5 dwarf provided by
Kenyon \& Hartmann (1995) we obtained an $E(B-V)=0.0$, so this method
implied no extinction to the star NTT 045251+3016.  Alternatively, we assumed the
\emph{H} band is unaffected by extinction and compared the calculated \emph{(V-H)} for
the primary with the \emph{(V-H)} for a K5 dwarf given in Kenyon \&
Hartmann (1995), finding a visual extinction of A$_{V}= 0.15 \pm 0.09$ mag
for the primary of NTT 045251+3016.  We adopt A$_{V}=0.15$ mag
throughout this paper.

Since there are temperature and color variations within
the K5 spectral class we define the range on these parameters by the
midpoints between the values given for a K5 type and the values for K3
and K7 types, respectively.  These upper and lower limits on
temperature or color are shown in Figures 4, 5, \& 6 as separate
points connected by a solid line.  This line represents the maximum
error range on temperature/color based on our spectral type
uncertainty, not a random error.  The smaller distance uncertainties
and photometric uncertainties are represented as error bars on each
point. Table 5 tabulates these stellar parameters and uncertainties
for the primary and secondary (Section 4.2) of NTT 045251+3016.

Two sets of Baraffe et al. (1998, BCAH98) tracks are plotted on $M_{V}
- (V-H)$ color-magnitude diagrams (CMDs) with NTT 045251+3016 (Figure
4).  Since the solar metallicity template provided the strongest
correlation when analyzing the optical spectra, we only test tracks
with solar metallicity.  BCAH98's use of the non-grey ``NextGen''
atmosphere models of Allard and Hauschildt (1997) allows their tracks
to be plotted directly on CMD diagrams.  The use of non-grey
atmosphere models is superior because grey atmospheres tend to
overestimate the luminosity and effective temperature of a star of a
given mass (BCAH98, and references therein).  Also, BCAH98 argue that
CMDs are superior to luminosity - $T_{eff}$ diagrams because the
latter usually depends upon empirically based color-$T_{eff}$ or
color-bolometric corrections which are derived from a stellar sample
with a range of metallicity, gravity, and age.

The PMS model that provides the closest agreement with our dynamical
primary mass is given by the BCAH98 tracks that use a general mixing
length parameter $\alpha$ = $l/H_{p} = 1.0$, [M/H] = 0.0, and Y =
0.275.  These tracks predict the mass of the primary to be between
1.15 - 1.20 M$_{\odot}$.  As seen in Figure 4, this mass range
deviates between 1.3 and 1.6 sigma from our dynamical primary mass
measurement.

The second set of PMS tracks shown in Figure 4 were designed by BCAH98
to reproduce the properties of the Sun at 4.61 Gyr. The primary
modification in these tracks is an increase in the mixing length
parameter, $\alpha$, along with a small change in the Helium abundance
($\alpha = 1.9$ and Y = 0.28). Using these tracks the derived mass
range of the primary is 1.05 M$_{\odot}$ to 1.15 M$_{\odot}$, which is
1.6 - 2.1 sigma from our dynamical primary mass. Thus while models
with this increased mixing length parameter do not provide as close
agreement with our measured dynamical mass as $\alpha = 1.0$ models,
they are nonetheless consistent with our measurements.  While the
difference between these two mixing length parameters is rather
inconsequential below $\sim 0.6$ M$_{\odot}$, it becomes important for
stars above this mass (BCAH98).

\begin{figure}[t]
\epsscale{0.55} 
\plotone{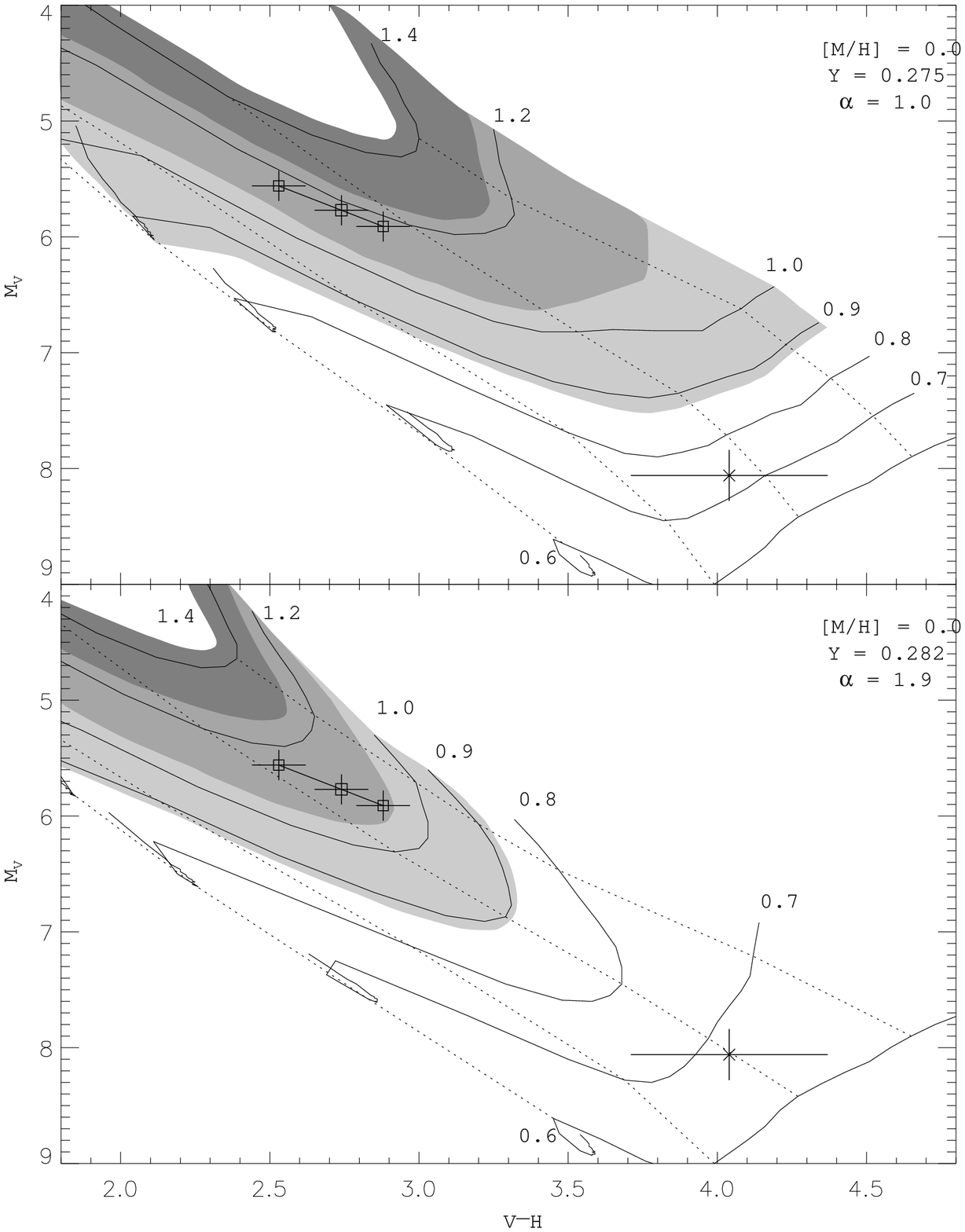}
\caption{Comparison of Baraffe et al. (1998) PMS tracks with NTT
045251+3016.  The metallicity, helium abundance, and mixing length
parameter of each set of tracks are shown.  The central open square on
each figure represents the intrinsic magnitude and color of NTT
045251+3016, corrected for extinction and distance as described in the
text. The other two open squares represent the range in these
corrections within a K5 spectral type.  The horizontal error bars
represent 1 sigma photometric uncertainties and the vertical error
bars represent both the combination of both photometric and distance
uncertainties. The solid lines are pre-main-sequence evolutionary
tracks for stellar masses as labeled. Regions with model masses within
1.0, 2.0, and 3.0 sigma from our dynamical primary mass measurement of
1.45 $\pm$ 0.19 M$_\odot$ are shaded from dark to light, respectively.
The dotted lines are isochrones starting at $3.16 \times 10^{6}$ yr
and increasing by a factor of $10^{0.5}$ yr. The secondary is plotted
with an X and its error bars represent 1.0 sigma observational errors.
(The 1.2 M$_{\odot}$ and 1.4 M$_{\odot}$ tracks were provided by
I. Baraffe via private communication.}
\label{fig:4}
\end{figure}

Recently, Palla \& Stahler (1999, PS99) calculated a set of PMS tracks
to model the star formation history of the Orion Nebula Cluster.
These tracks utilize the initial properties of the protostellar
environment to create a well-defined birthline in the H-R diagram
(PS99, and references therein).  This birthline assists in providing
more accurate age estimates to stars that have formed from the same
environment.

The PS99 tracks use a grey atmosphere approximation and employ the
opacities of Alexander \& Ferguson (1994) for low temperatures and the
OPAL opacities of Iglesis \& Rogers (1996) for high temperatures
($>10^{4}$K) with a composition X = 0.70, Y = 0.28 (PS99).  Convection
is treated using mixing length theory with $\alpha = 1.5$.  Figure 5
displays the PS99 tracks with NTT 045251+3016.  The mass range of 0.90
- 1.15 M$_{\odot}$ given to NTT 045251+3016 by the PS99 tracks
deviates from our measured primary mass by 1.6 - 2.9 sigma.

\begin{figure}[t]
\epsscale{1.0}
\plotone{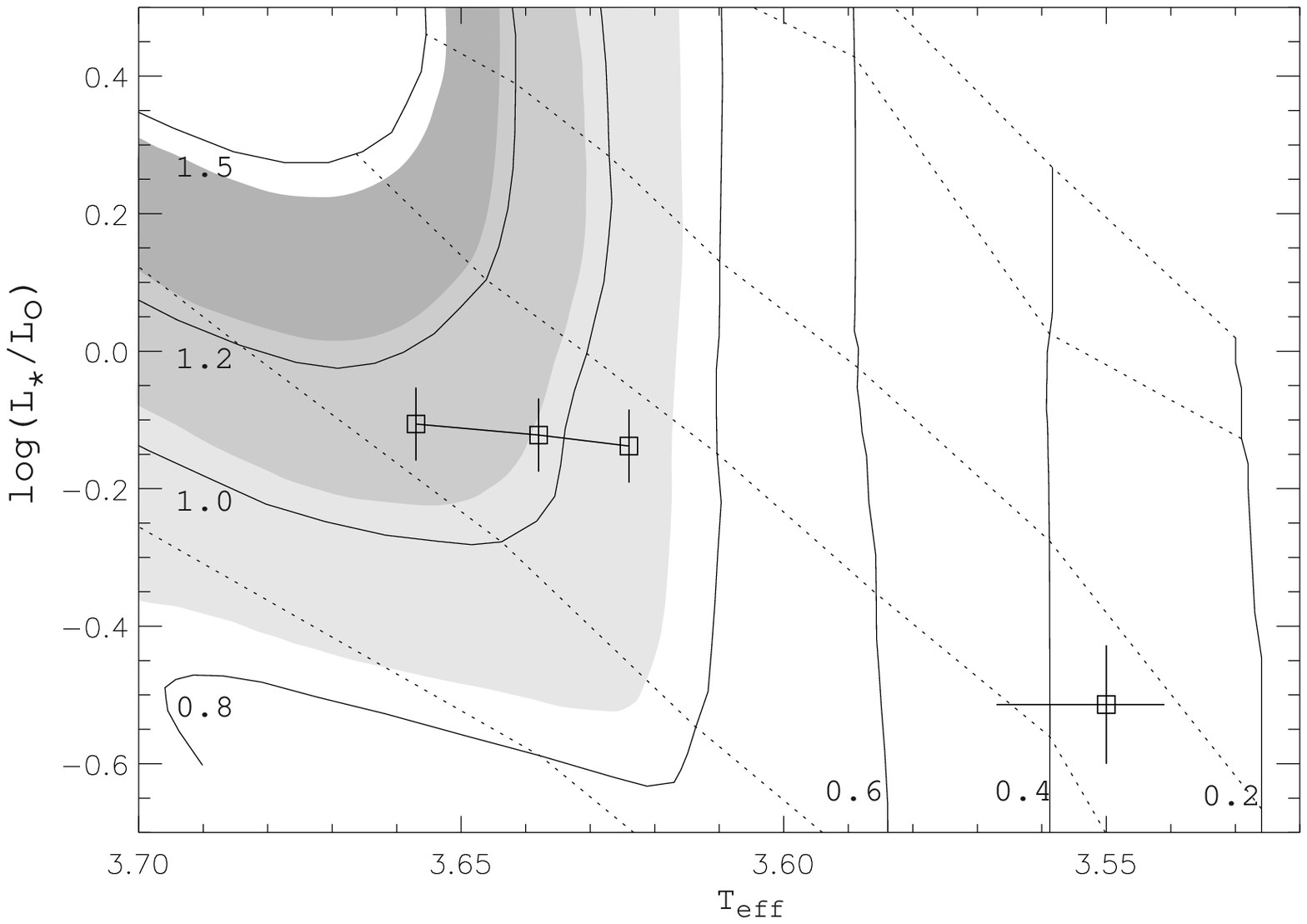}
\caption{Palla \& Stahler (1999) PMS tracks.  Data are as in Figure 4,
except presented in log(T$_{eff}$)-log(L$_{*}/$L$_{\odot}$) domain.
The dotted lines are isochrones that start at $1 \times 10^{5}$ yr and
increase in steps of $10^{0.5}$ yr.}
\label{fig:5}
\end{figure}

Our final comparison is with the PMS tracks of D'Antona \& Mazzitelli
(1997, DM97).  In 1994, D'Antona \& Mazzitelli released a set of PMS
tracks that did not use standard Mixing Length Theory (MLT) to model
the convection within the stellar envelope but instead treated the
extended envelope of a PMS star with a `multiple eddy' model called
the Full Spectrum of Turbulence (FST) model (Canuto \& Mazzitelli
1991).  In 1997, D'Antona \& Mazzitelli released a new set of PMS
tracks that used the updated FST convection model in Canuto, Goldman,
\& Mazzitelli (1996).

Figure 6 shows the DM97 tracks that provide the closest agreement with
our dynamical mass of NTT 045251+3016.  This set of tracks uses
Z=0.02, Y=0.28, and a deuterium mass fraction of $1 \times 10^{-5}$.
Nonetheless, the mass range given by DM97 for NTT 045251+3016 (0.60 -
0.87 M$_{\odot}$) deviates between 3.0 and 4.5 sigma from our
dynamical primary mass measurement.  From this result we conclude the
theoretical tracks provided by DM97 are not in agreement with our
observational results.

\begin{figure}[t]
\epsscale{1.0}
\plotone{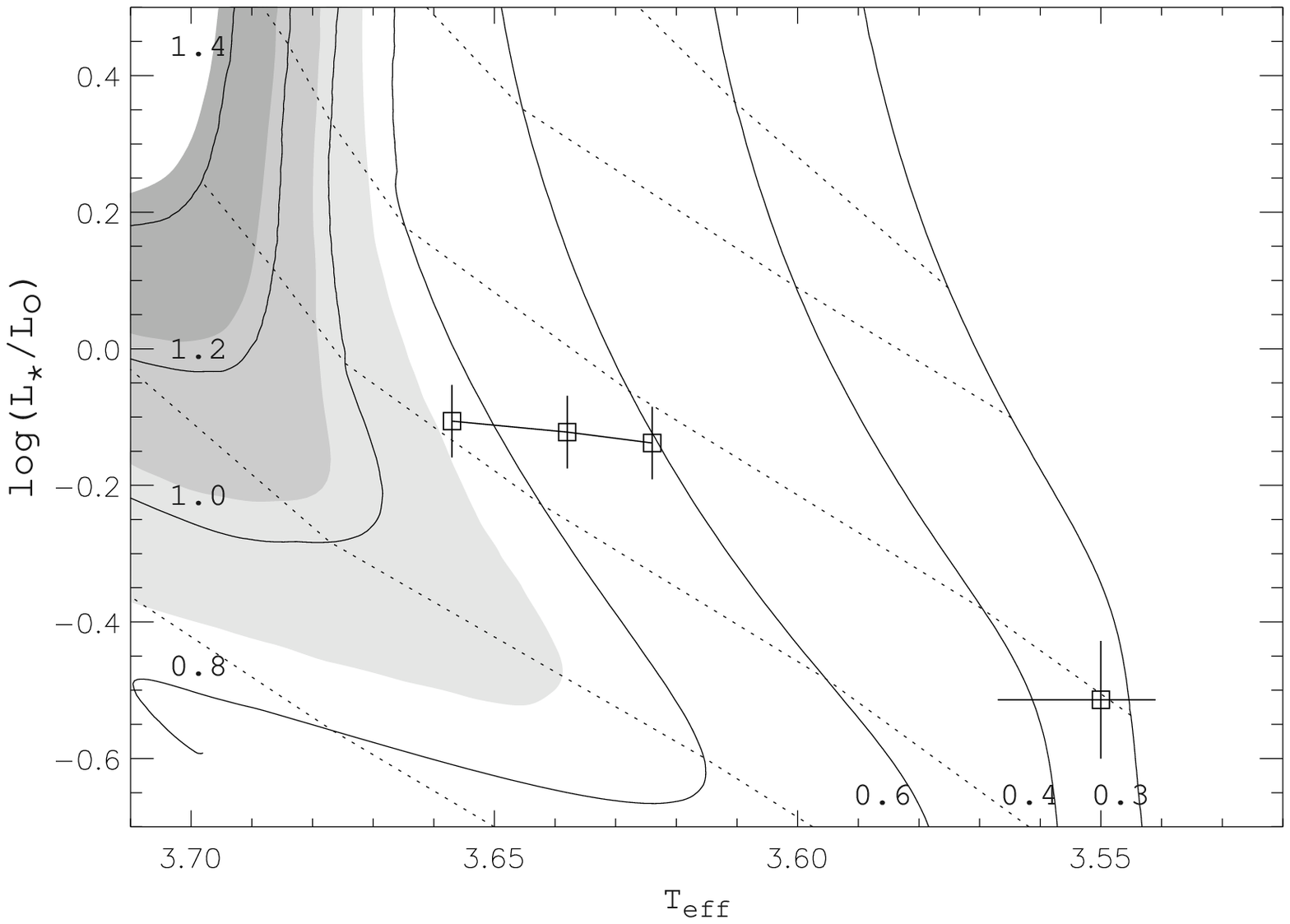}
\caption{D'Antona and Mazzitelli (1998) PMS tracks.  Data and curves
as in Figure 5.}
\label{fig:6}
\end{figure}

\subsection{The Secondary Star}
Given the flux ratio for the binary in two colors, \emph{V} and
\emph{H}, it is possible to use the photometry presented in Walter et
al. (1988) and the intrinsic colors of Kenyon \& Hartman (1995) to
deduce the spectral type of the secondary.  Using the extinction
coefficient calculated earlier ($A_{V}=0.15$ mag), we calculate $(V -
H)_{o} = 2.74 \pm 0.09$ for the primary star.  Applying the observed
flux ratios of $0.12 \pm 0.03$ and $0.4 \pm 0.1$ in the \emph{V} and
\emph{H} bands, respectively, we determine that for the secondary $(V
- H)_{o} = 4.04 \pm 0.33$. The uncertainty associated with the flux
ratio in the \emph{V} and \emph{H} bands is the main source of the
uncertainty in the observed color of the secondary.  This color
corresponds to an M2 spectral type plus or minus one subclass.  The
$T_{eff}$ was calculated from the $(V - H)$ color for the secondary by
interpolating in Kenyon \& Hartmann (1995).  The stellar parameters of
the secondary are tabulated in Table 5.

Knowledge of the mass ($0.81 \pm 0.09$ M$_{\odot}$), T$_{eff}$, and
luminosity of the secondary not only provides another constraint for
the mass calibration of the PMS models, but it also provides a test of
the isochrones for each PMS model, if we assume coevality of the
primary and secondary.  The issue of coevality in binary formation is
not yet well understood (e.g., Hartigan et al. (1994)), but the
relatively old age ($\sim 3-10$ Myr) of the NTT 045251+3016 pair makes
zero-point issues in age comparisons less of a concern. In addition,
we argue that the small physical separation of the stars in the binary
($4.75 \pm 0.33$ AU) suggests that both stars formed together rather
than through capture or exchange.

The BCAH98 PMS model that provides the best agreement with our
dynamical primary mass ($\alpha = 1.0$) also provides the best
agreement with our dynamical secondary mass.  The secondary mass of
0.73 M$_{\odot}$ predicted by the models is within 1.0 sigma of our
dynamical mass measurement. The isochrones are consistent with
coevality of the two stars in the binary and predict the age of the
components of the binary to be approximately $1.3x10^{7}$ yr.

The BCAH98 PMS model that uses a mixing length parameter $\alpha =
1.9$ predicts a mass of 0.67 M$_{\odot}$ for the secondary which is
within the 2.0 sigma range about our measured dynamical mass of the
secondary.  The primary star is estimated to be around $6x10^{6}$ yr
old while the secondary is about $10^{7}$ yr old.  However, the
stars are within 1.0 sigma of being coeval.  The larger observational
error on the secondary, compared to the error on the primary, is the
limiting factor in comparing the calibration of the isochrones.

Palla \& Stahler's PMS tracks (1999) give the mass of the secondary to
be 0.35 M$_{\odot}$, to be compared with our measured value for the
secondary mass of 0.81 M$_{\odot}$.  At this position among the PS99
tracks the predicted mass of the secondary depends only on
T$_{eff}$.  A 1.0 sigma variation on T$_{eff}$ corresponds to a change
in mass of 0.15 M$_{\odot}$. Even when combined with the error on the
dynamical mass of the secondary of 0.09 M$_{\odot}$, the mass predicted
by the PS99 tracks is 2.5 sigma from our measured dynamical mass.

Interestingly, the isochrones provided by PS99 suggest the largest relative age
difference in the two stars among the tracks tested in this paper.
Using the PS99 isochrones, the primary is over $6x10^{6}$ yr old while
the secondary is only $1.6x10^{6}$ yr old.  However, these age
determinations are only distinct between the 1 and 2 sigma confidence level.

The DM97 tracks also give a mass for the secondary of 0.34
M$_{\odot}$.  As with the PS99 tracks, the secondary's location with
respect to the DM97 tracks makes the predicted mass sensitive
primarily to changes in T$_{eff}$.  A 1.0 sigma change in T$_{eff}$
corresponds to a change in predicted mass of 0.1 M$_{\odot}$.
Combining this uncertainty with the error on the measured mass we find
that the secondary mass predicted by the DM97 tracks is at least 3
sigma from the dynamical mass.  The DM97 isochrones give the ages of
both the primary and secondary stars to be $1.8x10^{6}$ yr and
$10^{6}$ yr, respectively.  These ages are within 1 sigma of being
coeval.

\section{Conclusion}
In this paper we present dynamical masses for the components of the
binary NTT 045251+3016. These mass measurements were derived from
analysis of astrometric data from the Hubble Space Telescope Fine
Guidance Sensors, optical spectroscopic data from the digital
speedometers at CfA, and IR spectroscopic data from PHOENIX at KPNO.
Our measured values for the primary and secondary masses are 1.45
$\pm$ 0.19 M$_\odot$ and 0.81 $\pm$ 0.09 M$_\odot$, respectively, at a
distance of 145 $\pm$ 8 pc. The uncertainties in these mass measurements can
be readily reduced by more numerous measurements of the secondary radial velocity, which we encourage.

The measured primary and secondary masses are compared to predicted
masses of three sets of PMS tracks: Baraffe et al. (1998), Palla \&
Stahler (1999), and D'Antona \& Mazzitelli (1997).  The Baraffe et
al. (1998) tracks that use the ``Next Gen'' non-grey atmosphere models
of Allard and Hauschildt (1997) and a mixing length parameter $\alpha
= 1.0$ provide the closest agreement with our results. The masses
predicted by the models deviate between 1.3 and 1.6 sigma from our
measured primary mass and less than 1.0 sigma from our measured
secondary mass. The Baraffe et al. (1998) tracks with $\alpha = 1.9$
predict a primary mass that is between 1.6 and 2.1 sigma from our
measured primary mass, and a secondary mass that is within 2 sigma of
our dynamical value.  The PMS tracks of Palla \& Stahler (1999) give a
primary mass range that deviates 1.6 - 2.9 sigma from our dynamical
primary mass, and the predicted secondary mass is 2.5 sigma from our
dynamical value.  Finally, the values for the primary and secondary
masses provided by the tracks of D'Antona and Mazzitelli (1997), which
uses the Full Spectrum of Turbulance (FST) to model stellar convection
rather than standard Mixing Length Theory (MLT), deviate by more than
3.0 sigma from our measured dynamical masses.  We therefore conclude
that the Baraffe et al. (1998) and the Palla \& Stahler (1999) models
are consistant with our observations of NTT 045251+3016, while the
D'Antona \& Mazzitelli (1997) tracks are inconsistant at a significant
confidence level.

If we assume the binary system is coeval, we can use our observations
to constrain the relative accuracy of the isochrones provided by each
PMS model.  All three PMS models tested are consistant with coeval
formation of both components.  Better determination of the effective temperatures
of the component stars will be needed to provide a tighter
constraint on the PMS isochrones.

All of the PMS evolutionary tracks tested
in this paper underestimate the masses of both the primary and secondary stars. The secondary star lies amidst the Hayashi tracks, and
our observations demand cooler effective temperatures for the 0.8 Mo tracks. The same is true for the primary star when compared to the D'Antona and Mazzitelli
models. However, when compared to the Baraffe et al. and Palla \& Stahler tracks
the primary lies at the transition between the convective and radiative tracks.
For these models our observations of the primary star require both cooler temperatures and lower luminosities. This need for cooler effective temperatures from the evolutionary models is similar to the recent findings of Simon et al. (2001). Interestingly,  D'Antona, Ventura, and
Mazzitelli (2000) argue that the effective temperatures predicted by present PMS
evolutionary models are actually upper limits, because magnetic
fields, which are not included in any of the models tested here,
act to reduce effective temperature.

\section{Acknowledgements}

We are indebted to the Space Telescope Science Institute, and
especially the FGS team, for superb technical support throughout this
program and for funding support. We also thank the many CfA observers
who have obtained observations over the 15 years of this program and
E. Goldberg for helping with the TODCOR analysis. Finally, the
generous contribution of software from G. Torres, analyses and
expertise from C. Dolan, and additional evolutionary tracks from
I. Baraffe, were invaluable.  The work of Tsevi Mazeh was supported by
US-Israel Binational Science Foundation grant no. 97-00460, and by the
Israel Science Foundation.  The work of Lisa Prato and Michal Simon
was supported in part by US NSF grant 98-19694.

% References :

\begin{deluxetable}{cccccccc}
\tablewidth{0pt}
\tablenum{1}
\tablecaption{Relative astrometric measurements of NTT 045251+3016.}
\tablehead{\colhead{HJD} & \colhead{$\Delta m_{X}$(mag)} & \colhead{$\Delta m_{Y}$(mag)} & \colhead{PA\tablenotemark{1} ($\degr$)} & \colhead{$\rho$($\arcsec$)} & \colhead{PA($\degr$)(O-C)} & \colhead{$\rho$($\arcsec$)(O-C)} & \colhead{Phase}}
  \startdata
    2449822.0  &2.0  &2.2  &187  &0.046   &\phn\phs7.20  &\phs0.0052  &2.277 \\
    2450038.0  &2.1  &2.6  &178  &0.048   &\phn\phs4.75  &\phs0.0041  &2.363 \\
    2450098.0  &2.4  &2.2  &170  &0.048   &\phn$-$1.53 &\phs0.0041  &2.386 \\
    2450161.0  &1.9  &2.2  &169  &0.042   &\phn$-$0.68  &$-$0.0016     &2.412 \\
    2450317.0  &2.7  &2.5  &163  &0.041   &\phn$-$1.99  &$-$0.0005     &2.473 \\
    2450372.0  &2.2  &2.5  &160  &0.036   &\phn$-$3.17  &$-$0.0043     &2.495 \\
    2450480.0  &2.4  &2.4  &158  &0.039   &\phn$-$1.27  &\phs0.0015  &2.537 \\
    2450538.0  &2.3  &2.6  &153  &0.037   &\phn$-$3.88  &\phs0.0013  &2.560 \\
    2450669.0  &2.1  &2.2  &149  &0.031   &\phn$-$1.36  &\phs0.0000  &2.612 \\
    2450705.0  &1.7  &2.7  &146  &0.029   &\phn$-$2.19  &$-$0.0005  &2.626 \\
    2450767.0  &2.1  &2.2  &138  &0.016   &\phn$-$5.89  &$-$0.0110     &2.651 \\
    2450882.0  &2.6  &1.8  &124  &0.014   &\phn$-$9.21  &$-$0.0081     &2.697 \\
    2450913.0  &2.2  &2.4  &121  &0.023   &\phn$-$8.48  &\phs0.0022  &2.709 \\
    2451046.0  &2.5  &2.1  &120  &0.018   &\phs12.93      &\phs0.0020  &2.762
\enddata
\tablenotetext{1}{The Position Angle (PA) is measured North through East.}
\end{deluxetable}

\begin{deluxetable}{cccc}
\tablecolumns{4}
\tablewidth{0pt}
\tablenum{2}
\tablecaption{Optical radial-velocity measurements for NTT 045251+3016}
\tablehead{\colhead{HJD} & \colhead{$v_{1}$ \kms} & \colhead{$v_{1}$(O-C) \kms} & \colhead{Phase} }
\startdata
    2446421.7465  &\phn4.52   &$-$0.47    &0.930 \\
    2446428.6909  &\phn5.26   &\phs0.39    &0.933 \\
    2446451.6493  &\phn3.38   &$-$1.12    &0.942 \\
    2446728.8724  &11.84   &\phs0.45    &1.052 \\
    2446775.7779  &13.31   &\phs0.04    &1.071 \\
    2446804.6601  &13.92   &$-$0.34    &1.082 \\
    2447045.0073  &18.00   &$-$0.35    &1.177 \\
    2447075.8236  &18.86   &\phs0.30    &1.189 \\
    2447080.8803  &17.56   &$-$1.03    &1.191 \\
    2447127.8694  &19.15   &\phs0.34    &1.210 \\
    2447138.7852  &18.99   &\phs0.14    &1.214 \\
    2447157.5923  &19.36   &\phs0.44    &1.222 \\
    2447192.6571  &18.86   &$-$0.15    &1.236 \\
    2447198.6334  &20.13   &\phs1.11    &1.238 \\
    2447427.9813  &19.53   &\phs0.51    &1.329 \\
    2447492.8084  &19.37   &\phs0.47    &1.355 \\
    2447546.5820  &19.49   &\phs0.71    &1.376 \\
    2447576.6672  &18.96   &\phs0.26    &1.388 \\
    2447791.0071  &16.73   &$-$1.21    &1.473 \\
    2447818.8032  &18.62   &\phs0.80    &1.484 \\
    2447837.7737  &16.72   &$-$1.01    &1.491 \\
    2447868.9265  &17.52   &$-$0.07    &1.504 \\
    2447899.5659  &17.73   &\phs0.29    &1.516 \\
    2447928.6599  &15.84   &$-$1.45    &1.527 \\
    2447957.6175  &17.40   &\phs0.26    &1.539 \\
    2447965.6088  &17.31   &\phs0.22    &1.542 \\
    2448168.9077  &15.89   &\phs0.06    &1.622 \\
    2448194.9668  &15.10   &$-$0.54    &1.633 \\
    2448284.7375  &14.39   &$-$0.55    &1.668 \\
    2448635.7208  &11.43   &\phs0.46    &1.807 \\
    2448669.5847  &11.13   &\phs0.70    &1.821 \\
    2448675.6738  &11.39   &\phs1.06    &1.823 \\
    2448697.6311  &10.25   &\phs0.28    &1.832 \\
    2448875.9712  &\phn6.05   &$-$0.38    &1.902 \\
    2448901.9111  &\phn5.34   &$-$0.54    &1.913 \\ 
    2448910.8049  &\phn6.41   &\phs0.72    &1.916 \\
    2448931.9191  &\phn6.84   &\phs1.57    &1.925 \\
    2448970.8280  &\phn4.23   &$-$0.36    &1.940 \\
    2448988.7716  &\phn5.38   &\phs1.04    &1.947 \\
    2449030.7093  &\phn3.75   &$-$0.31    &1.964 \\
    2449056.6040  &\phn3.41   &$-$0.74    &1.974 \\
    2449235.0189  &11.21   &\phs0.67    &2.045 \\
    2449258.9979  &12.14   &\phs0.52    &2.054 \\
    2449290.8841  &13.23   &\phs0.32    &2.067 \\
    2449316.8181  &13.11   &$-$0.73    &2.077 \\
    2449318.8322  &13.72   &$-$0.19    &2.078 \\
    2449379.6724  &14.56   &$-$1.09    &2.102 \\
    2449652.0326  &18.03   &$-$0.78    &2.210 \\
    2449706.8549  &17.98   &$-$1.00    &2.231 \\
    2450000.9524  &19.19   &\phs0.25    &2.348 \\
    2450029.9190  &18.76   &$-$0.12    &2.359 \\
    2450087.7501  &18.47   &$-$0.26    &2.382 \\
    2450173.5382  &19.51   &\phs1.04    &2.416 \\
    2450771.7774  &16.51   &\phs1.26    &2.653 \\ 
    2450771.7982  &14.68   &$-$0.57    &2.653 \\
    2450799.9570  &14.96   &$-$0.06    &2.664 \\
    2451563.5879  &\phn2.21   &$-$1.85    &2.967 \\
    2451570.4936  &\phn5.53   &\phs1.45    &2.970
  \enddata
\end{deluxetable}

\begin{deluxetable}{cccccc}
\tablewidth{0pt}
\tablenum{3}
\tablecaption{IR radial-velocity measurements for NTT 045251+3016}
\tablehead{\colhead{HJD} & \colhead{$v_{1}$ \kms} & \colhead{$v_{1}$(O-C) \kms} & \colhead{$v_{2}$ \kms} & \colhead{$v_{2}$(O-C) \kms} &\colhead{Phase}}
\startdata
    2451529.7500  &\phn4.1   &$-$0.08    &33.6  &\phs0.70  &2.953 \\
    2451530.7500  &\phn3.4   &$-$0.77    &31.6  &$-$1.04   &2.954
  \enddata
\end{deluxetable}

\begin{deluxetable}{lr}
\tablewidth{0pt}
\tablenum{4} 
\tablecaption{Orbital Elements for NTT 045251+3016
(Astrometric-Spectroscopic Solution)}
\tablehead{\colhead{} & \colhead{}}
  \startdata
    Period (yr) &$6.913 \pm 0.033$\\
    $\emph{a}$ ($''$) &$0.0328 \pm 0.0013$\\
    $\emph{e}$  &$0.457 \pm 0.017$\\
    $\emph{i}$ ($^\circ$)     &$113.8 \pm 3.4$\\
    $\omega$ ($^\circ$)  &$216.7 \pm 2.8$\\
    $\Omega_{2000}$ ($^\circ$)  &$179.5 \pm 2.7$\\
    T$_{o}$              &$1993.369 \pm 0.042$ \\
    $\emph{K}_1$ (km s$^{-1}$)              &$7.53 \pm 0.16$ \\
    $\emph{K}_2$ (km s$^{-1}$)              &$13.52 \pm 0.67$ \\
    \medskip
    $\gamma$ (km s$^{-1}$)             &$14.35 \pm 0.11$ \\
    $\emph{a}$ (AU) &$4.75 \pm 0.33$\\
    \medskip
    distance (pc) &$144.8 \pm 8.3$\\
    $\emph{M}_1$ (M$_\odot$) & $1.45 \pm 0.19$\\
    $\emph{M}_2$ (M$_\odot$) & $0.81 \pm 0.09 $\\
    $\emph{M}_1$ + $\emph{M}_2$ (M$_\odot$) & $2.26 \pm 0.21$\\
    \smallskip
    $\emph{q} \equiv$ $\emph{M}_2$/$\emph{M}_1$ & $0.56 \pm 0.03$
  \enddata
\end{deluxetable}  

\begin{deluxetable}{cccccc}
\tablewidth{0pt}
\tablenum{5} 
\tablecaption{} 
\tablehead{\colhead{\phs} & \colhead{\phs} & \colhead{log($T_{eff}$)} & \colhead{log($L_{*}/L_{\odot})$} & \colhead{$(V-H)_{o}$} & \colhead{$M_{V}$}}
\startdata 

     Primary &Central point &$3.638$ &$-0.122 \pm 0.053$ &$2.74 \pm 0.09$ &$5.77 \pm 0.13$ \\
     \phs &lower limit &$3.624$ &$-0.138 \pm 0.053$ &$2.88 \pm 0.09$ &$5.91 \pm 0.13$ \\
     \phs &upper limit &$3.657$ &$-0.106 \pm 0.053$ &$2.53 \pm 0.09$ &$5.56 \pm 0.13$ \\
     Secondary &\phs &$3.550^{+0.017}_{-0.009}$ &$-0.514 \pm 0.086$ &$4.04 \pm 0.33$ &$8.06 \pm 0.22$ \\
  \enddata
\end{deluxetable}  

\end{document}